\documentclass[11pt,a4paper]{article}
 
\pdfoutput=1 

\usepackage{graphicx}
\usepackage{epstopdf}
\usepackage{jcappub}
\usepackage{amsmath,amsthm,latexsym,amssymb,amsfonts,epsfig}
\usepackage{fontenc,dsfont,layout}
\usepackage{hyperref}
\usepackage{psfrag}
\usepackage[cp1250]{inputenc}
\usepackage[usenames,dvipsnames]{xcolor}

\newcommand{\bea}{\begin{eqnarray}}
\newcommand{\eea}{\end{eqnarray}}
\newcommand{\beq}{\begin{eqnarray}}
\newcommand{\eeq}{\end{eqnarray}}

\newcommand{\GeV}{\,\text{GeV}}

\numberwithin{equation}{section}

\title{Hill-climbing dark inflation}

\abstract{Within the framework of the scalar-tensor theory we consider a hill-climbing inflation, in which the effective Planck mass increases in time. We obtain the Einstein frame potential with infinitely long and flat plateau as we approach towards the strong coupling regime, together with a run-away vacuum in the GR limit of the theory. The inflation ends with the scalar field rolling down towards infinity, which at the effective level indicates the massless scalar field domination in the Universe. In this scheme we assume that the inflaton is a dark particle, which has no couplings to the Standard Model degrees of freedom (other than the gravitational ones). We discuss the gravitational reheating of the Universe together with its implications on the predictions of the model, including possible amplification of primordial gravitational waves. Our model for the first time realizes explicitly the enhancement of the primordial gravitational waves in the dark inflation scenario.}


\author[a,b]{Micha{\l} Artymowski}
\author[a]{Zygmunt Lalak}
\author[c]{Kin-ya Oda}
\affiliation[a]{Institute of Theoretical Physics, Faculty of Physics, University of Warsaw, ul. Pasteura 5, 02-093 Warsaw, Poland}
\affiliation[b]{Institute of Physics, Jagiellonian University, {\L}ojasiewicza 11, 30-348 Krak{\'o}w, Poland}
\affiliation[c]{Department of Physics, Osaka University, Osaka 560-0043, Japan}
\emailAdd{Michal.Artymowski@fuw.edu.pl}
\emailAdd{Zygmunt.Lalak@fuw.edu.pl}
\emailAdd{Odakin@phys.sci.osaka-u.ac.jp}

\begin{document} 

\begin{flushright}
OU-HET/975
\end{flushright}

\maketitle

\section{Introduction}

The cosmic inflation \cite{Starobinsky:1980te} is a hypothetical era in the evolution of the early Universe characterised by the accelerated expansion of space; see Ref.~\cite{Lyth:1998xn} for a review. It can be responsible for diluting unwanted relics~\cite{Sato:1980yn} and for solving problems of horizon and curvature \cite{Guth:1980zm}, as well as for the generation of primordial inhomogeneities, which are the seeds of the present large scale structure of the Universe. Inflation seems to be perfectly consistent with observational data \cite{Array:2015xqh} and regardless of some issues \cite{Ijjas:2013vea} (see Ref.~\cite{Linde:2017pwt} for an opposite point of view) the inflationary paradigm have become a canonical theoretical framework for the analysis of the evolution of the early Universe.

Inflation is usually generated by scalar fields with flat potentials or by some modification of general relativity (GR). In the latter case, one of the most prevailing theories that generate primordial accelerated expansion are the scalar-tensor theories \cite{Kallosh:2013tua,Kallosh:2014laa,Giudice:2014toa,Kallosh:2014rha,Artymowski:2016dlz}, for which the Ricci scalar is directly coupled to a function $F$ of a Jordan frame scalar field $\phi_\text{J}$. Usually one assumes that $F$ tends to be much bigger than 1 during inflation, which leads effectively to weaker gravity and inflation. The other approach is to consider $F \ll 1$, which pushes the model towards the strong coupling regime for the gravity. As shown in Refs.~\cite{Jinno:2017jxc,Jinno:2017lun,Kawana:2017rgu}, while $F$ is growing, the field $\phi_\text{J}$ may go uphill on its potential and still generate inflation, which is called the hill-climbing inflation.

In this paper we explore the idea that the inflation could be generated by a hill-climbing field from the dark sector, in which fields (possibly accounting for dark energy and dark matter) do not have direct couplings with the Standard Model (SM) fields other than the gravitational ones. We assume that it is protected by some symmetries or any other mechanism that would prevent it from decay into the SM degrees of freedom. Such an inflaton would not have any coupling besides the gravitational one to other fields. We will show that this idea is consistent with the gravitational reheating scenario \cite{Artymowski:2017pua} and it can be fully consistent with the current observational data. Gravitational reheating is based on the observation that even if the inflaton is a dark field, one can still reheat the Universe via the gravitational coupling between the dark and SM sectors. This coupling cannot be fully screened and it inevitably produces scalar particles at the end of inflation. The aim of this paper is to present the concrete physical realization of dark inflation within the scalar-tensor theory framework.

Throughout the paper we will use the convention $8\pi G = M_\text{p}^{-2}$, where $M_{p} = 2.435\times 10^{18}$ GeV is the reduced Planck mass.

The structure of the paper goes as follows: In Sec.~\ref{sec:HCinflation} we consider the background evolution of the hill-climbing inflation. In Sec.~\ref{sec:reheating} we discuss the gravitational reheating in this theory. In Sec.~\ref{sec:inhomogeneities} we present the features of the primordial inhomogeneities produced during inflation, together with the constrains from the observational data.

\section{Hill-climbing inflation} \label{sec:HCinflation}

\begin{figure}
\includegraphics[height=6cm]{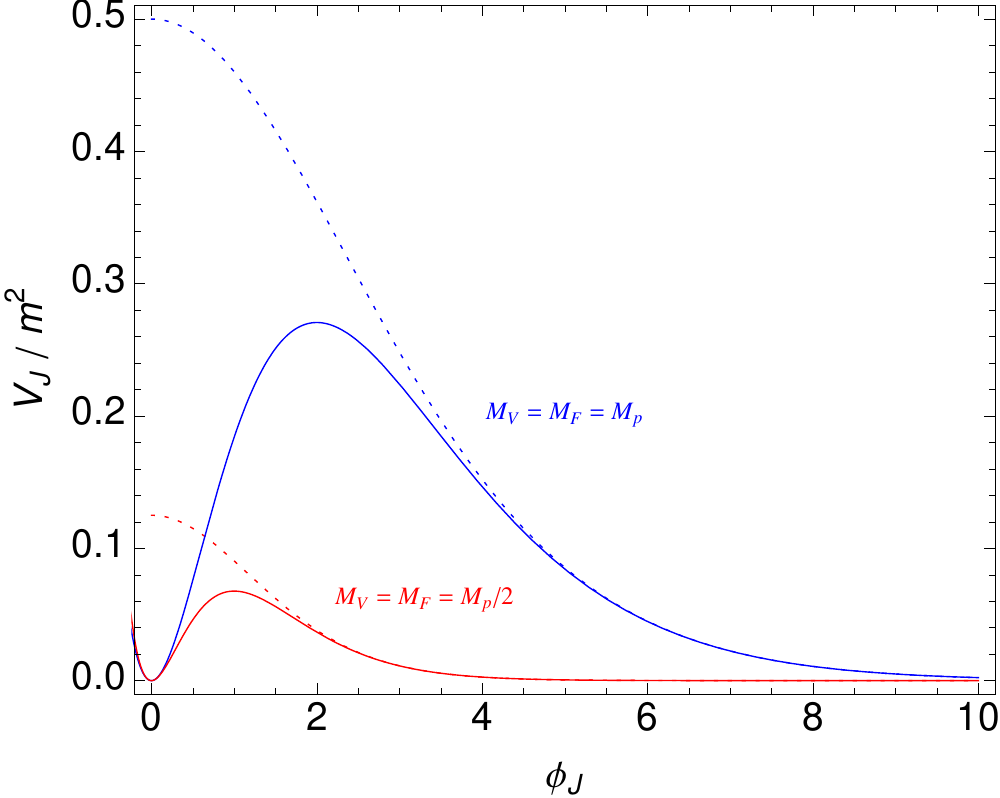}  
\hspace{0.5cm}
\includegraphics[height=6cm]{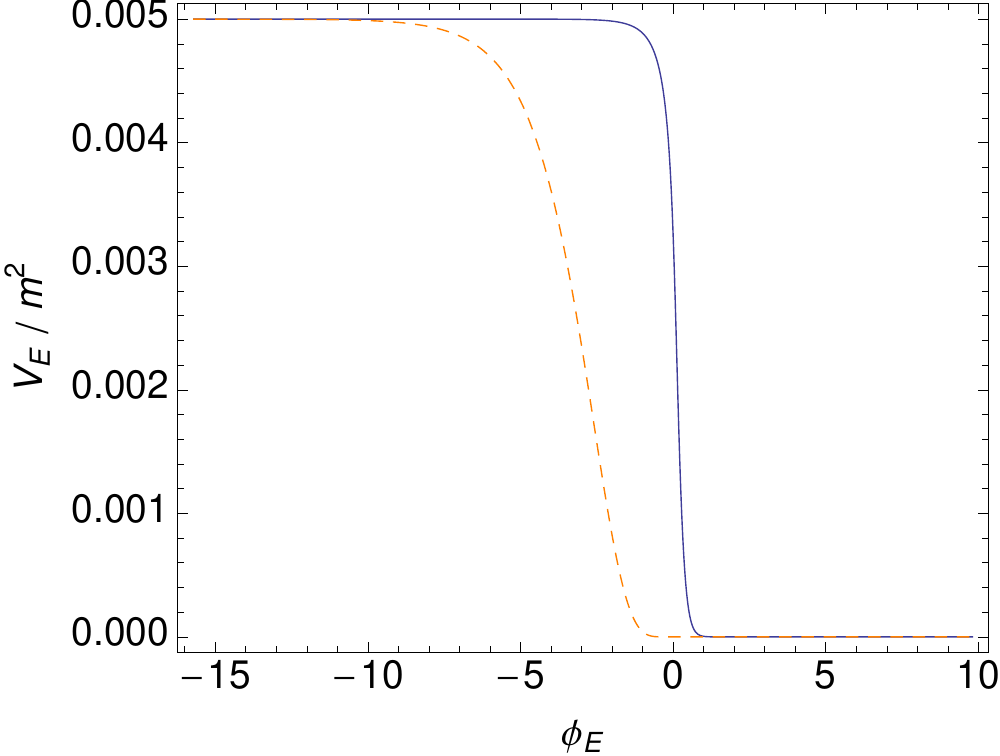}  
\caption{Left panel: the Jordan frame potential (solid lines) for 2 different values of $M_V$. The barrier between the vacuum in $\phi_\text{J} = 0$ and a runaway vacuum in $\phi_\text{J} \to \infty$ appears at $\phi_{\text{J}_{\max}} = 2 M_V$, which gives $V_\text{J}(\phi_{\text{J}_{\max}}) = 2 m^2M_V ^2e^{-2}$. The Einstein frame potentials as a function of $\phi_\text{J}$ are plotted in dotted lines. We have assumed $M_V = M_F$. Right panel: The Einstein frame potential as a function of an Einstein frame field. Blue (dashed orange) lines represent the $M_F = M_V = 0.1 M_\text{p}$ ($M_F=10M_V = 0.1 M_\text{p}$) case. One can see the inflationary plateau and the steep slope, which will be the source of the kinaton domination.}
\label{fig:Jordan framepotential}
\end{figure}

We start from the action of a scalar-tensor theory
\begin{equation}
S = \int d^4 x\sqrt{-g}\left[\frac{M_\text{p}^2}{2}F(\phi_\text{J})R-\frac{1}{2}\left(\partial\phi_\text{J}\right)^2-V_\text{J}(\phi_\text{J})\right] \, ,
\end{equation}
where $\phi_\text{J}$ and $V_\text{J}$ are the Jordan frame (Jordan frame) inflaton field and the potential, respectively, and $F$ is a function that defines the non-minimal coupling to gravity. Let us consider the following forms of $F$ and $V_\text{J}$:
\begin{align}
V_\text{J}
	&=	\frac{1}{2}m^2\phi_\text{J}^2 e^{-\phi_\text{J}/M_V},	&
F 	&= 	1-e^{-\phi_\text{J}/M_F},\label{eq:Jordan framepotential}
\end{align}
where $M_V$, $m$, and $M_F$ have a dimension of mass. For $\phi_\text{J} = 0$ one obtains $F = 0$, which is the strong coupling limit for the gravity.  For $\phi_\text{J} < 0$ one finds $F < 0$, which is the repulsive gravity regime. Therefore we will restrict our analysis to $\phi_\text{J}>0$. In the $\phi_\text{J} \gg M_F$ limit one obtains $F \to 1$, which restores GR. The Jordan frame potential has been presented in the left panel of Fig.~\ref{fig:Jordan framepotential}. It has two vacua---one in $\phi_\text{J} = 0$ and a run-away vacuum in $\phi_\text{J} \to \infty$. 

The model presented by us is rather phenomenological. We have chosen forms of the potential and a function of non-minimal coupling in order to obtain an inflationary plateau and a kinaton phase after inflation. Nevertheless the model could be embedded in more fundamental theories. For instance, we note that the exponential runaway potential naturally arises from string theory; see e.g.\ Ref.~\cite{Hamada:2015ria,Lalak:2005hr}.

Alike any other scalar-tensor theory, our model can be expressed in the Einstein frame, which is defined by the Weyl rescaling
\begin{equation}
g_{\text{E}\mu\nu}=F g_{\mu\nu}\, .
\end{equation}
Then, one obtains the Einstein frame action of the form of 
\begin{equation}
S= \int d^4 x \sqrt{-g_\text{E}}\left[ \frac{1}{2}R_\text{E} - \frac{1}{2}\left( \partial\phi_\text{E} \right)^2 - V_\text{E}(\phi_\text{E}) \right]\, ,
\end{equation}
where $R_\text{E}$ is the Einstein frame Ricci scalar. Note that in the Einstein frame one restores the action of a scalar field with a minimal coupling to gravity. The $\phi_\text{E}$ in the Einstein frame inflaton reads
\begin{equation}
\phi_\text{E} = \int d\phi_\text{J}\sqrt{\frac{1}{F} + \frac{3}{2}M_\text{p}^2\left(\frac{F_{,\phi_\text{J}}}{F}\right)^2} \, , \label{eq:phi}
\end{equation}
where $F_{,\phi_\text{J}} = \frac{dF}{d\phi_\text{J}}$. The $V_\text{E}$ is the Einstein frame potential defined by
\begin{equation}
V_\text{E} = \left. \frac{V_\text{J}(\phi_\text{J})}{F^2(\phi_\text{J})} \right|_{\phi_\text{J} = \phi_\text{J}(\phi_\text{E})} \, .
\end{equation}
For models with $F\gg 1$, such as e.g.\ $\xi$-attractors, one obtains a flat plateau of $V_\text{E}$ for big values of $V_\text{J}$, i.e. when $V_\text{J} \to F^2$. In our model we investigate a different approach. We consider small values of $F$ and $V_\text{J}$ in order to obtain a constant value of the Einstein frame potential in the small field limit, defined by $\phi_\text{J} \ll M_F,M_V$. Then one finds
\begin{equation}
F \simeq \frac{\phi_\text{J}}{M_F}\left(1-\frac{1}{2}\frac{\phi_\text{J}}{M_F}\right) \, ,\qquad V_\text{J} \simeq \frac{1}{2}m^2\phi_\text{J}^2\left(1-\frac{\phi_\text{J}}{M_V}\right) \, ,
\end{equation}
which gives the following Einstein frame potential
\begin{equation}
V_\text{E} \simeq \frac{1}{2}m^2 M_F^2\left(1-\frac{M_F-M_V}{M_F M_V}\phi_\text{J}\right)\, . \label{eq:V1}
\end{equation}
Here we assume the following:
\begin{itemize}
\item The starting value of $\phi_\text{J}$ is small and positive.
\item The following relation holds:
\begin{equation}
M_F > M_V
\end{equation}
so that the field increase its value over time in order not to evolve towards $F=0$, which is the strong coupling limit of the theory.
\item Initially, $\dot{\phi}_\text{J}$ is not to negative so that the universe does not go into the repulsive gravity regime.
\end{itemize}
In the same small $\phi_\text{J}$ limit one finds
\begin{equation}
\left(\frac{F_{,\phi_\text{J}}}{F}\right)^2 F \simeq \frac{M_F}{\phi_\text{J}}\frac{M_\text{p}^2}{M_F^2} \, .
\end{equation}
Therefore for $\phi_\text{J} M_F \ll M_\text{p}^2$, one finds $(F_{,\phi_\text{J}}/F)^2 \gg 1/F$, which is the strong coupling limit of the theory. In such a case one can estimate $\phi_\text{E}$ to be
\begin{equation}
\phi_\text{E} \simeq \sqrt{\frac{3}{2}}\log F \qquad \Rightarrow \qquad \phi_\text{J} \simeq M_F e^{\sqrt{2/3}\, \phi_\text{E}} \, . \label{eq:Einstein framefield1}
\end{equation}
From Eq.~\eqref{eq:V1} one can see that the Einstein frame potential obtains a constant value in the small field limit. In the Einstein frame, $\phi_\text{J} = 0$, i.e.\ the strong coupling limit of the theory, corresponds to $\phi_\text{E} = -\infty$. The Einstein frame field is therefore always infinitely far from the region, where gravity becomes non-perturbative. The form of the Jordan frame potential and the chosen form of $F$ are fitted to satisfy the $V_\text{E} \to const$ limit for $\phi_\text{J} \to 0$.

From Eq.~(\ref{eq:phi}), we see that $\phi_\text{E} \to \infty$ corresponds to $\phi_\text{J} \to \infty$, which is the GR limit of the theory. From Eqs.~(\ref{eq:V1}) and (\ref{eq:Einstein framefield1}) one finds
\begin{equation}
V_\text{E}(\phi_\text{E}) \simeq \frac{1}{2}m^2 M_F^2\left(1-\frac{M_F-M_V}{M_V}e^{\sqrt{2/3} \, \phi_\text{E}}\right) \, , \label{eq:V2}
\end{equation}
which is a Starobinsky-like model.

It would be interesting to investigate loop corrections to model, especially in the $\phi_\text{J} \to 0$ limit. Note that the loop corrections to the Einstein frame potential may spoil the perfect flatness of the plateau in the strong coupling limit. Nevertheless, this should not influence the results presented in this paper. We are considering the evolution of the field at the last (i.e. observable) stage of inflation, which is rather far from the strong coupling limit (corresponding to $\phi_\text{E} \to - \infty$). In fact the loop corrections may play a useful role in the context of avoiding the eternal inflation, which happens when the field starts its evolution at the extremely flat part of the potential. This issue will be studied in our further work.\footnote{Note that the issue of fermionic loop corrections in a general scalar-tensor theory has already been analyzed in the Ref. \cite{Hamada:2016onh}. However, in our case the only loop corrections that may occur must come from the graviton loops since the inflaton is a dark field. This issue was partially analyzed in the Ref.~\cite{Oda:2015sma}.}

Note that the case of $M_V = M_F$ also gives inflationary solution. Then Eq.~(\ref{eq:V1}) takes form of
\begin{equation}
V_\text{E} \simeq \frac{1}{2}m^2M_F^2 \left(1-\frac{1}{12}\left(\frac{\phi_\text{J}}{M_F}\right)^2\right) = \frac{1}{2}m^2M_F^2\left(1-\frac{1}{12}e^{2\sqrt{2/3} \, \phi_\text{E}}\right) \, . \label{eq:Vequal}
\end{equation}

Another interesting case is the $M_V \ll \phi_\text{E} \ll M_F$ limit. It can be obtained for a strong hierarchy between mass scales $M_F$ and $M_V$. In such a case Eq.~(\ref{eq:Einstein framefield1}) still holds, but one cannot consider $\phi_\text{J}/M_V$ to be a small parameter. The Einstein frame potential takes the form
\begin{equation}
V_\text{E} = \frac{1}{2}m^2M_F^2 \exp\left(-\frac{M_F}{M_V}e^{\sqrt{2/3} \, \phi_\text{E}}\right) \, .
\end{equation}
This limit corresponds to the dark blue line in Figs. \ref{fig:RNs} and \ref{fig:RNsFull}.

The Einstein frame slow-roll parameters are defined by
\begin{align}
\epsilon 
	&= \frac{M_\text{p}^2}{2}\left(\frac{V_{\text{E},\phi_\text{E}}}{V_\text{E}}\right)^2 \, , &
\eta 
	&= M_\text{p}^2\frac{V_{\text{E},\phi_\text{E}\phi_\text{E}}}{V_\text{E}} \, . \label{eq:srparameters}
\end{align}
In the $\phi_\text{J} \ll M_F,M_V$ limit one finds
\begin{align}
\epsilon 
	&\simeq \frac{(M_F-M_V)^2 \phi_\text{J} ^2}{3M_F^2 M_V^2} \, , &
\eta 
	&\simeq -\frac{2}{3}\frac{(M_F-M_V)\phi_\text{J}}{M_F M_V} \, , \label{eq:srparameters1}
\end{align}
which generally shows that in the small field limit one should expect $\epsilon \ll |\eta|$. 

In the case of $M_F = M_V$ one obtains
\begin{align}
\epsilon 
	&\simeq \frac{\phi_\text{J} ^4}{108 M_F^4} \, , & 
\eta 
	&\simeq -\frac{2 \phi_\text{J} ^2}{9 M_F^2} \, . \label{eq:srparameters2}
\end{align}
Since we have assumed that $\phi_\text{J} \ll M_F,M_V$ one should expect rather small values of $\epsilon$ for both, $M_F\gg M_V$ and $M_F \simeq M_V$. Nevertheless, from Eqs.~(\ref{eq:srparameters1}) and (\ref{eq:srparameters2}) one can see that the slow-roll parameters should be much smaller in the $M_V \to M_F$ case. Due to the normalization of inhomogeneities the tensor-to-scalar ratio $r = 16\epsilon$ is proportional to the value of the potential at the pivot scale. Thus, for $M_V \to M_F$, the scale of inflation should become smaller than in the $M_F \gg M_V$ case. The Jordan frame and Einstein frame potentials have been plotted in Fig.~\ref{fig:Jordan framepotential}. One can see that the Einstein frame potential obtains a steep slope shortly after the plateau.

\begin{figure}
\centering
\includegraphics[height=4.5cm]{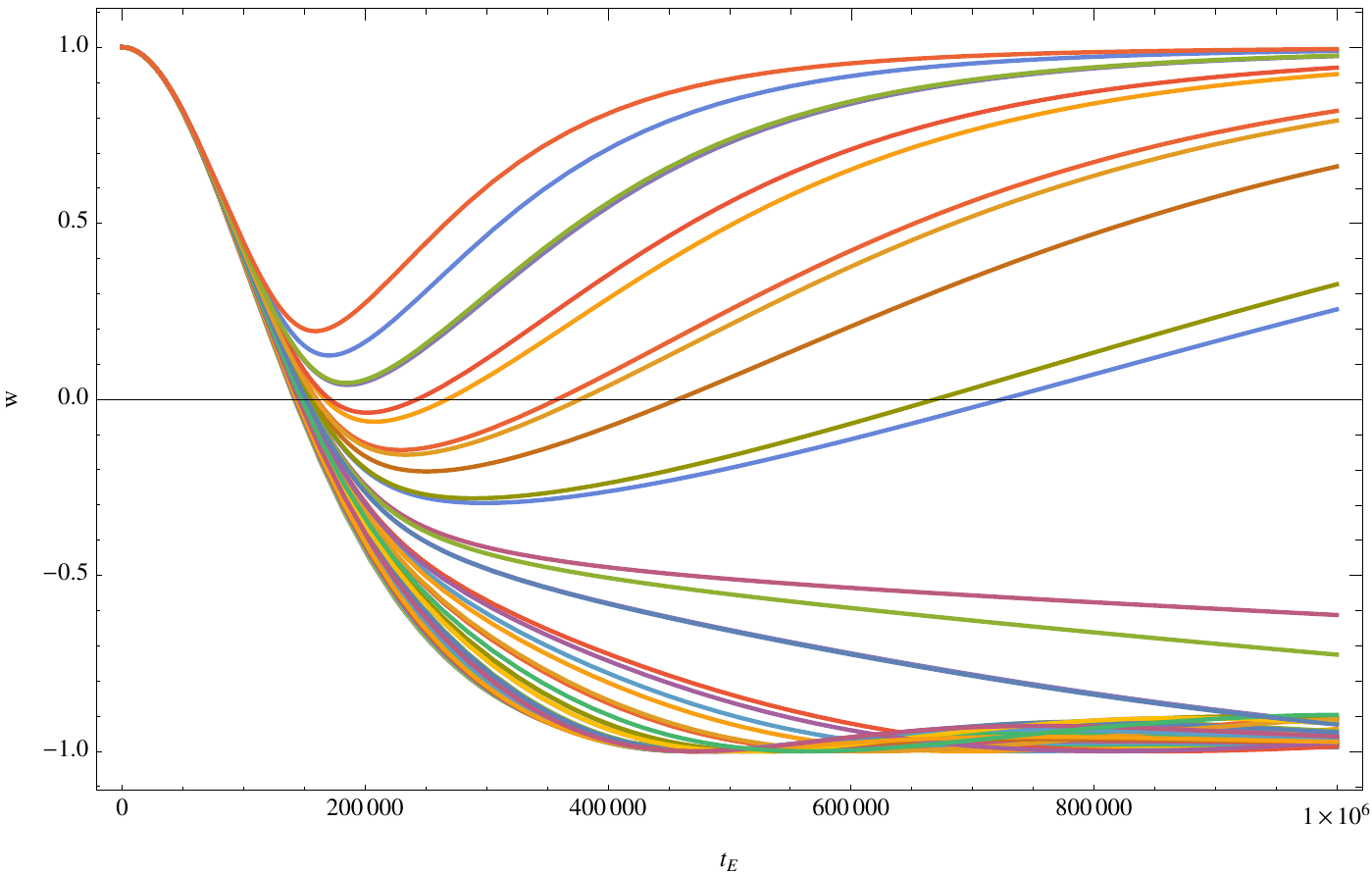}  
\caption{ The evolution of the barotropic parameter of the inflaton as a function of the Einstein frame time. We have assumed initial domination of the kinetic term in order to show that for many trajectories this shouldn't spoil the inflation. After reaching $w=-1$ the inflaton starts deviating more from the perfect de-Sitter, which finishes with $w=1$, which is the equation of state for a massless scalar field. }
\label{fig:barotropic}
\end{figure}

Another useful limit to consider is the other side of a local maximum of the potential. While the field is rolling down towards the runaway vacuum one finds $\phi_\text{J} \gg M_F,M_V$. This can be approximated by the $\phi_\text{J} \to \infty$ limit,\footnote{Indeed, the field tends to obtain arbitrarily big values after inflation. This should not be an issue, since its energy density decreases exponentially with the field value of the inflaton.}
which gives 
\begin{align}
\epsilon &\to \frac{M_\text{p}^2}{2 M_V^2} \, , & \eta &\to \frac{M_\text{p}^2}{M_V^2} \, ,
\end{align}
which is perfectly consistent with the predictions of the Einstein frame exponential potential. For $M_V > M_\text{p}$ one could obtain another phase of an accelerated expansion. Nevertheless, we are rather interested in the $M_V < M_\text{p}$ regime, which provides decelerated expansion of the post-inflationary Universe. In particular the post-inflationary evolution of an inflaton may strongly resemble a massless scalar field with the equation of state $\rho \simeq p$ (as shown in Fig.~\ref{fig:barotropic}). In Sec.~\ref{sec:reheating} we will show how such a steep post-inflationary potential and rapid decrease of inflaton's energy density may be used in the context of gravitational reheating.

\section{Reheating} \label{sec:reheating}
As mentioned, inflation generates quasi-de Sitter expansion of the Universe, with very slow redshift of the inflaton's energy density. This leads to a strong suppression of energy densities of all other fields and matter components, including radiation. Since the radiation domination is required at the scale of Big Bang nucleosynthesis (BBN)~\cite{Cyburt:2015mya}, one needs to include a mechanism of the reheating of the Universe after inflation. The reheating is usually equivalent to some particle production mechanism, which in most cases comes from direct couplings of the inflation to the SM  fields. 

It is possible that the dark matter as well as the dark energy belongs to what-we-call the dark sector (coupled to others only via gravitation), given especially the recent fast development of the dark matter experiments, which put severer and severer upper bound on the dark matter coupling to the SM fields; see Ref.~\cite{Aprile:2018dbl} for the latest. 
Under this circumstance, it is tempting to consider that the dark sector plays crucial role in the inflationary era or in other periods of the evolution of the Universe before the BBN. 
In this paper we assume that the inflaton itself is a dark field. In such a case the only mechanism of reheating is the gravitational particle production \cite{Artymowski:2017pua,Ford:1986sy,Kunimitsu:2012xx,Dimopoulos:2017zvq,Peebles:1998qn,deHaro:2017nui,AresteSalo:2017lkv}, which is restricted to the production of scalar fields. Note that fermions and vectors may also be produced during the transition between two gravitational vacua. Nevertheless, energy densities related to them are too small to significantly contribute to the reheating of the Universe \cite{Parker:1969au,Parker:1971pt}.

We emphasize that this mechanism should always produce particles at the end of inflation. It is usually neglected in the analysis of the post-inflationary Universe, since direct couplings to other fields provide much more efficient mechanism of the production of relativistic degrees of freedom. The fact that the gravitational reheating is always present at the end of inflation is an additional theoretical motivation for our work. We do not need to assume any new physics that would describe interactions between SM fields and the inflaton in order to explain the radiation domination. Furthermore, we avoid possible loop corrections (caused by interactions of the inflaton with other fields) to the inflationary potential that could in principle change the flatness of the potential and therefore change the predictions of given inflationary model \cite{Artymowski:2016dlz}. 

As shown in \cite{Artymowski:2017pua} a rapid transition between the de-Sitter spacetime and a Universe filled with a perfect fluid with a constant barotropic parameter $w$ generates the following energy density of radiation\footnote{
Though Eq.~\eqref{eq:radiation} has been rigorously derived, it might also be naively interpreted as $\rho\sim T^4$, with $T$ being the de Sitter temperature $T\sim H_\text{inf}$.
}
\begin{equation}
\rho_\text{r} \simeq H_{\rm inf}^4\frac{9 \textsf{N} (1-6\xi)^2 (1+w)^2}{128\pi^2} \left( \frac{a_\text{end}}{a} \right)^4 \, , \label{eq:radiation} 
\end{equation}
where $H_{\rm inf}$ is a Hubble parameter during inflation,
 $a_\text{end}$ is a scale factor at the end of inflation, $\textsf{N}$ is the number of scalar species produced gravitationally, and $\xi$ is a value of a direct coupling of a produced scalar field $\chi$ to the Ricci scalar of the form of $\xi \chi^2 R$. Thus, the gravitational particle production can be strongly suppressed by the non-minimal coupling close to the conformal value. The $(1-6\xi)^2$ factor was introduced by Ford in \cite{Ford:1986sy}. The analysis was restricted to rather small values of $\xi$ and therefore it's a non-trivial question how big values of $\xi$ amplify gravitational particle production. The value of $\textsf{N}$ can be ass small as $\textsf{N}=4$ in the case of SM or much bigger,\footnote{
For a temperature much higher than the electroweak scale $\sim100\GeV$, we may count the number of the SM degrees of freedom by two for each (nearly) massless vector, whose mass is negligible compared to the temperature, and by four for the Higgs, including the Nambu-Goldstone modes.
}
like $\textsf{N}\sim 100$ in the case of SUSY. As expected, produced radiation redshifts like $a^{-4}$. 

In fact $H_{\rm inf}$ can be also estimated as a Hubble parameter at the end of inflation, since for many models consistent with the data, such as Starobinsky inflation \cite{Starobinsky:1980te}, (critical) Higgs inflation \cite{Salopek:1988qh,Bezrukov:2007ep,Hamada:2014iga,Bezrukov:2014bra,Hamada:2014wna}, $\alpha$-attractors \cite{Kallosh:2013yoa,Artymowski:2016pjz,Dimopoulos:2016yep}, $\xi$-attractors and other scalar-tensor theories \cite{Kallosh:2014laa,Giudice:2014toa,Artymowski:2016ikw} etc., scale of inflationary plateau is very close to the scale of the end of inflation. This approximation works perfectly well in our model, since $H_{\rm inf}$ is limited from above by the scale of the plateau. 

Note that at the end of inflation one finds 
\begin{equation}
\rho_{\rm inf} \sim M_\text{p}^2H_{\rm inf}^2 \gg \rho_\text{r} \sim H_{\rm inf}^4 \, .
\end{equation}
This inequality is trivially satisfied since $H_{\rm inf}$ is limited from above by around $8 \times 10^{13}\GeV\sim 3\times 10^{-5}M_\text{p}$, which comes from the upper bound on the tensor-to-scalar ratio $r<0.09$ (95\% C.L.)~\cite{Array:2015xqh}. That is, the gravitational reheating is so inefficient that one obtains a domination of the inflaton field at the end of inflation. On the other hand we know from the BBN constrains~\cite{Cyburt:2015mya} that the Universe needs to be dominated by the SM radiation at the MeV scale. We have assumed that the inflaton is a dark particle and therefore it cannot dissipate into radiation by a direct coupling. Therefore, the only chance to obtain radiation domination is to have a quickly red-shifting of inflaton's energy density in between inflation and the BBN era. 

From the continuity equation for the perfect fluid with a constant barotropic parameter, one finds
\begin{equation}
\rho_i \propto a^{-3(1+w_i)} \, ,
\end{equation}
where $\rho_i$ is an energy density associated with given $w_i$. Radiation redshifts like $a^{-4}$, which corresponds to $w=1/3$. Therefore in order to obtain radiation domination in late times  one needs $w > 1/3$ for the inflaton's energy density after inflation. This may be obtained for e.g.\ oscillating scalar field \cite{Ford:1986sy} or massless scalar field \cite{Kunimitsu:2012xx, Kobayashi:2010cm, ArmendarizPicon:1999rj, Garriga:1999vw,Helmer:2006tz}. The latter case may be realized also for a field with a very steep potential, for which the kinetic term dominates over the potential one \cite{Dimopoulos:2017zvq}. Our potential has been designed in this way, to naturally satisfy the $w > 1/3$ condition. As mentioned in Sec. \ref{sec:HCinflation} one finds $w=1$ for any $M_V < M_\text{p}$, which is rather natural assumption to make. For $w=1$ the energy density of the inflaton redshifts much faster than radiation and therefore the gravitational reheating can be applied in our model.

\section{Primordial inhomogeneities and implications of dark inflation on cosmological evolution} \label{sec:inhomogeneities}

All variables given in this section will be defined in the Einstein frame. For the minimally coupled scalar field one finds within the slow-roll approximation the following values of the tensor-to-scalar ratio and of the scalar spectral index $n_s$:
\begin{align}
r	&= 16 \epsilon, &
n_s 
	&= 1 - 6\epsilon + 2\eta .
\end{align}
In order to compare the results with the PLANCK/BICEP data one needs to set $r$ and $n_s$ at the pivot scale.
In the standard inflationary scenario the scale of reheating is determined by the exact forms of couplings between the inflaton and other fields. The couplings are often independent of the model of inflation, and the scale of reheating may vary from the MeV scale up to the GUT scale, which is the maximal scale of inflation. In the case of dark inflation the reheating (understood as the beginning the the radiation domination era) is set by the scale of inflation and therefore one can reduce the uncertainty on the scale of reheating and therefore on $N_\star$. In \cite{Artymowski:2017pua} it has been proven that for the gravitational reheating one finds
\begin{equation}
N_\star \simeq 64.82 + \frac{1}{4} \ln\left( \frac{128 \pi^2}{\textsf{N}(1-6\xi)^2 (1+w)^2} \right) \, .
\end{equation}
In \cite{Artymowski:2017pua} we have presented $N_\star$ for wide range of $\textsf{N}$ and $w$ with a conclusion that in the case of dark inflation the pivot scale should leave the horizon around $N_\star \simeq 65$. The result can be additionally decreased by 1 or less even if the scale of inflation at the pivot scale is much bigger than the one at the end of inflation.

\begin{figure}[h!]
\includegraphics[height=5.5cm]{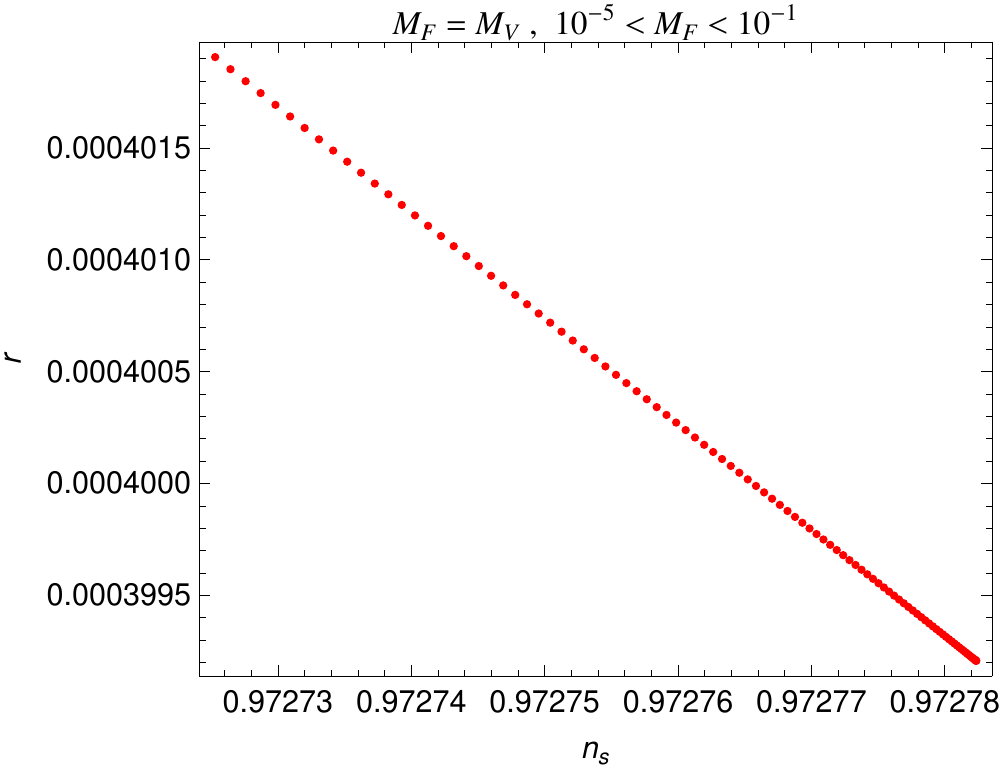}  
\hspace{0.5cm}
\includegraphics[height=5.5cm]{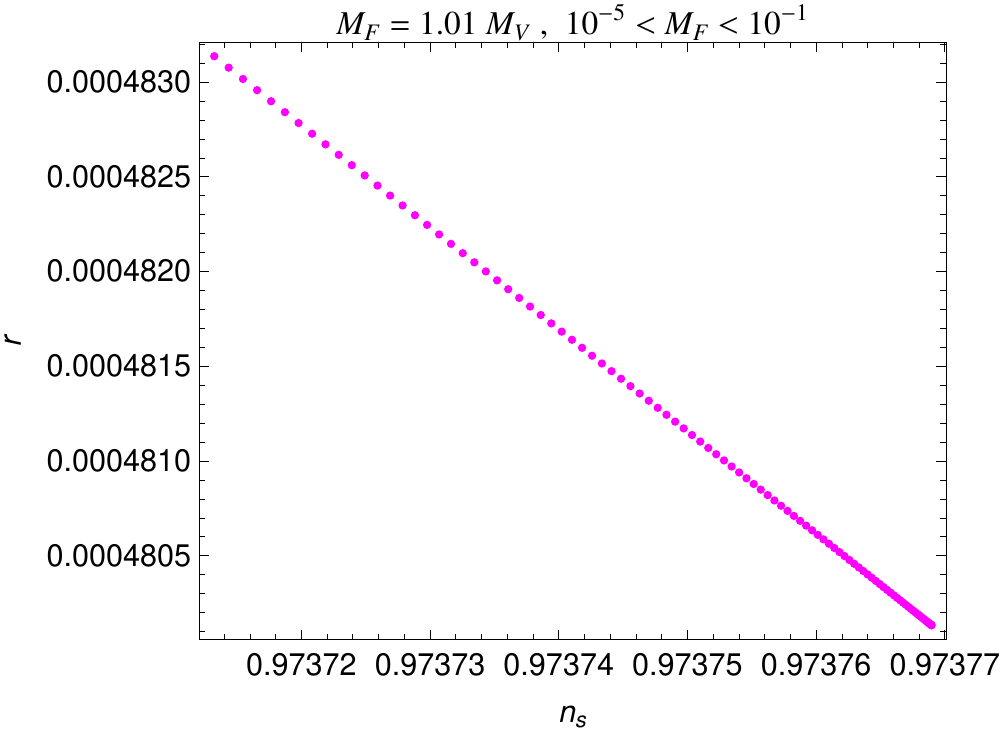}\\
\vspace{0.3cm}

\includegraphics[height=5.5cm]{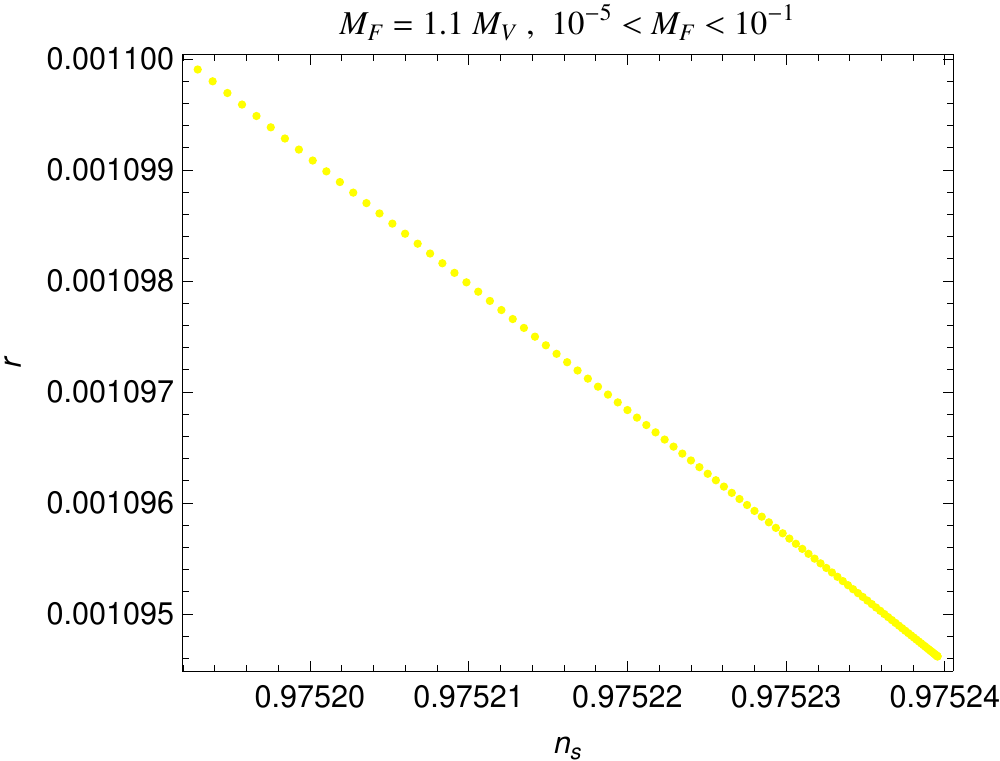}  
\hspace{0.5cm}
\includegraphics[height=5.5cm]{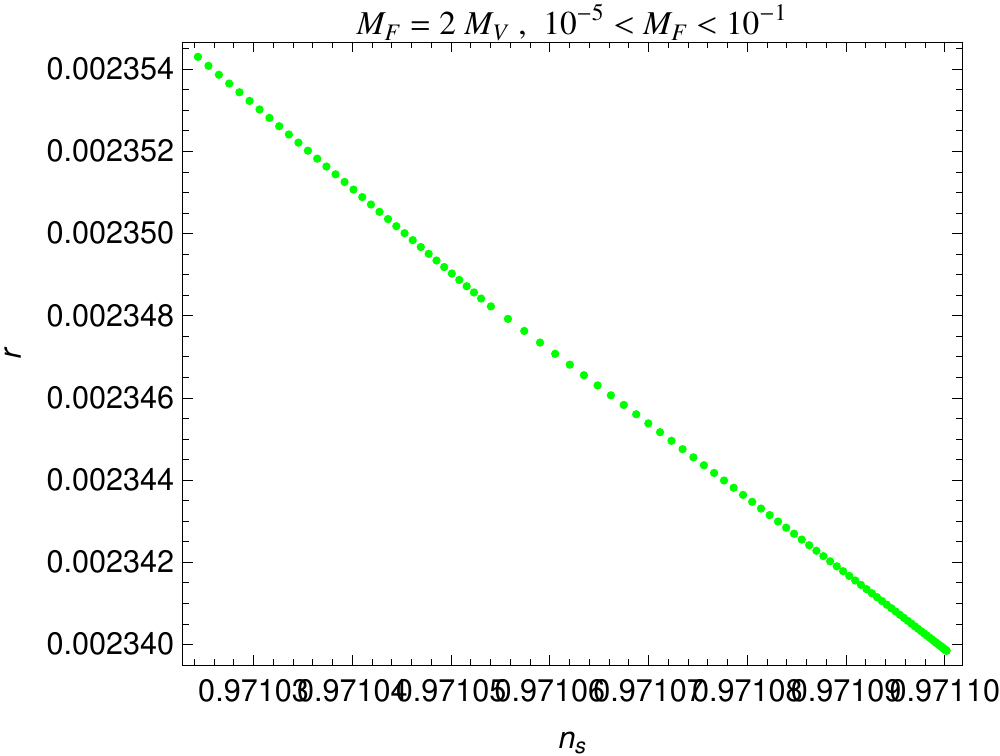}\\
\vspace{0.3cm}

\includegraphics[height=5.5cm]{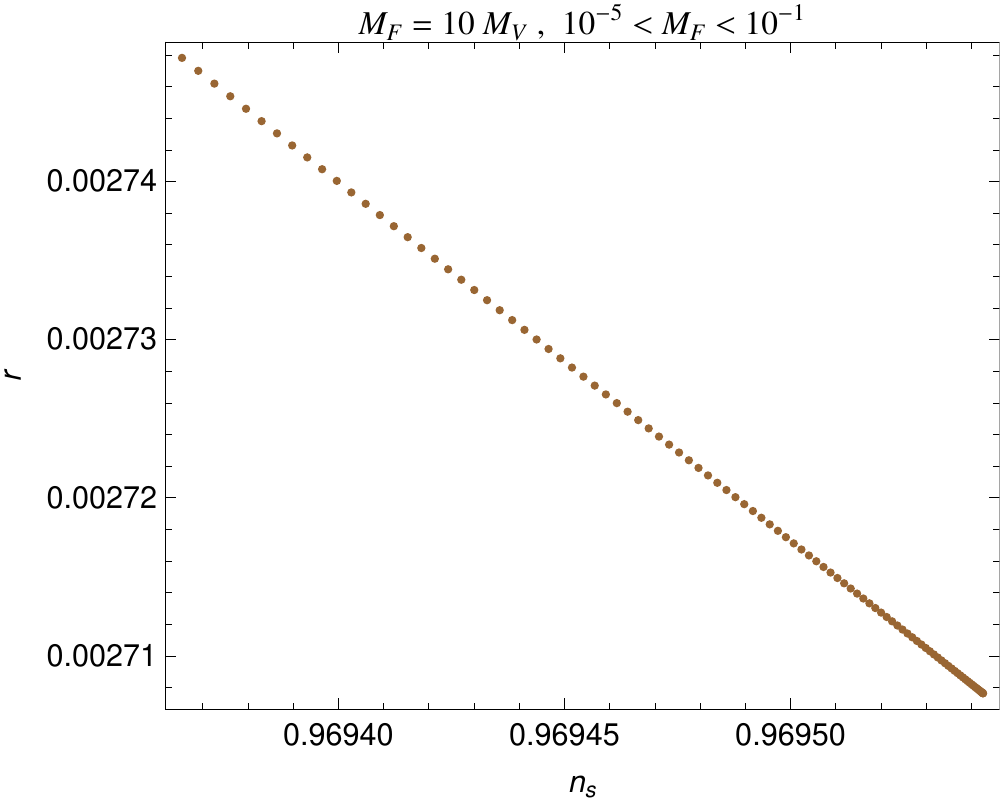}  
\hspace{1cm}
\includegraphics[height=5.5cm]{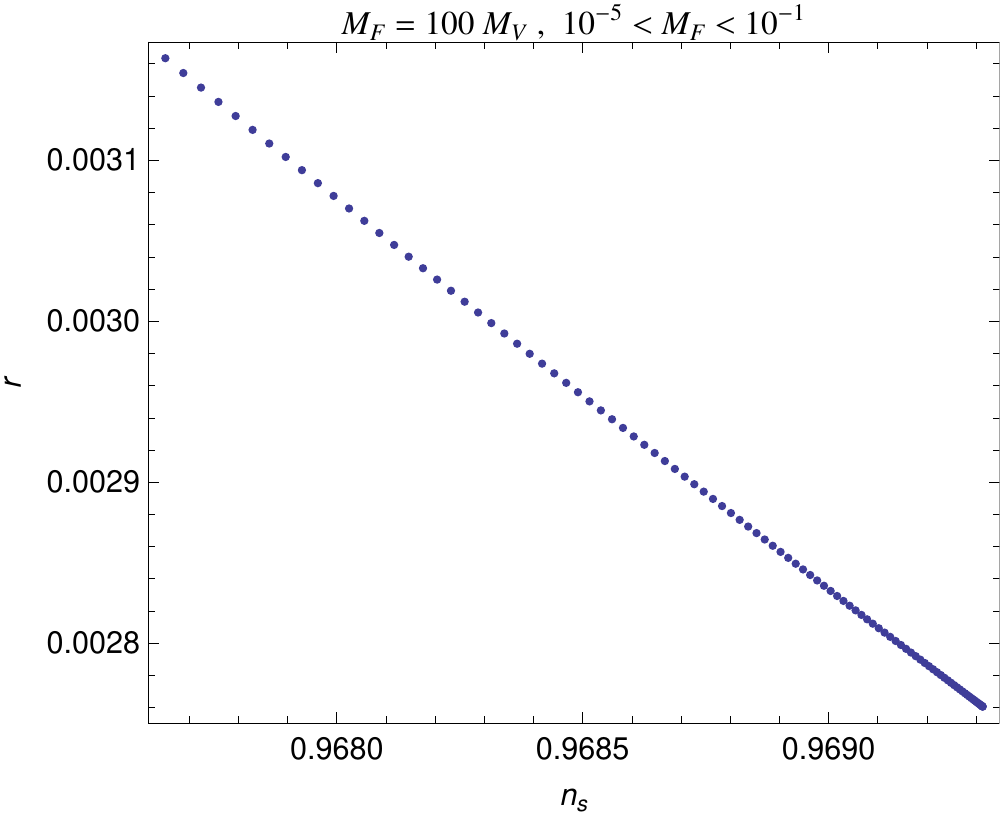}\\
\caption{ All figures present the tensor-to-scalar ratio as a function of a spectral index for $N_\star = 65$. All results are perfectly consistent with the PLANCK/BICEP data. As expected, $r$ is the smallest for $M_V = M_F$.}
\label{fig:RNs}
\end{figure}

\begin{figure}[h!]
\centering
\includegraphics[height=4.5cm]{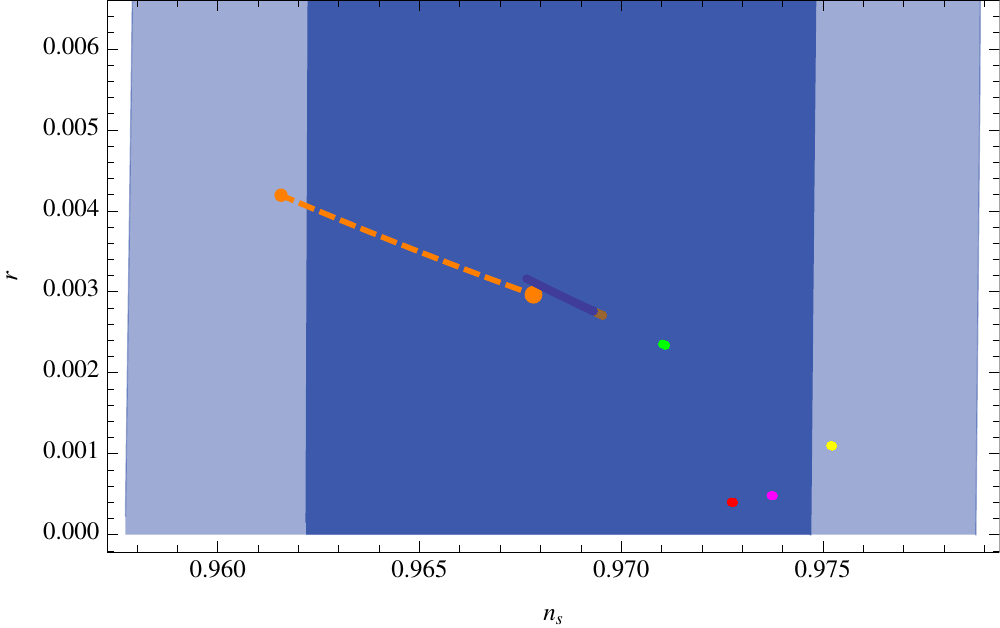}  
\hspace{0.5cm}
\includegraphics[height=4.5cm]{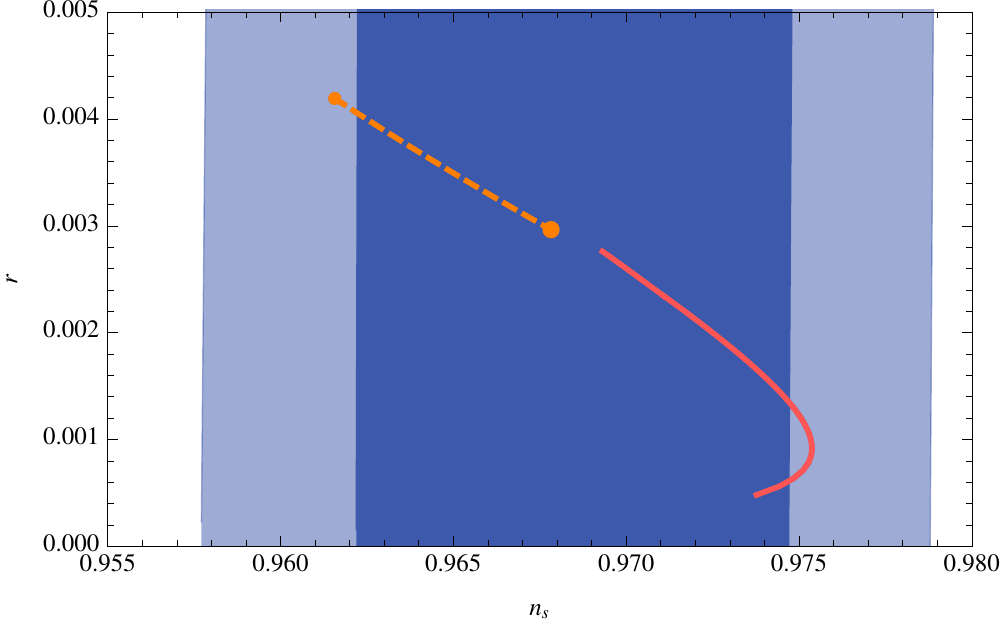}  
\caption{ Left panel: The results from all of the plots presented in the fig. \ref{fig:RNs}, together with the 1$\sigma$ and 2$\sigma$ limits from the PLANCK/BICEP. Again, $M_F \in [10^{-5},10^{-1}]$ and $M_F/M_V=\{1, 1.01, 1.1, 2, 10, 100\}$ for red, pink, yellow, green, brown and dark blue lines respectively. One can see that for $M_F \gg M_V$ the results tend to coincide with the Starobisnky inflation. Besides the $M_F = 100 M_V$ case all of the lines from the fig. \ref{fig:RNs} looks like points, due to the fact that in those cases the results tend to depend very weakly on $M_F$. Right Panel: Results for $M_V = 10^{-3}$ and $M_F \in [M_V,100M_V]$.}
\label{fig:RNsFull}
\end{figure}

In Figs. \ref{fig:RNs} and \ref{fig:RNsFull} we present the results at the $(n_s,r)$ plane for $N_{\star} = 65$. We consider a wade range of parameters $M_F$ and $M_V$. As expected, the result seems to coincide with the Starobisnky inflation for $M_F \gg M_V$.  Note that in most cases the result depend on the ratio between $M_F$ and $M_V$ rather than on $M_F$ itself. In Fig.~\ref{fig:RNsFull}, the result for the Starobinsky inflation has also been presented for, as usual, $N_\star \in (50,60)$ by the (orange) dashed line.

The primordial metric perturbations produced during inflation consist of scalar and tensor components. The latter ones have flat inflationary power spectrum, which is usually too small to be measured by future gravitational waves experiments. Nevertheless, as shown in Ref. \cite{Artymowski:2017pua}, the contribution of the primordial gravitational waves to the total energy density of the Universe can be enhanced if the gravitational reheating in very inefficient, which is equivalent to the condition $\textsf{N} (1-6\xi)^2 \ll 1$, where the non-minimal coupling of produced scalars are close to the conformal value $\xi = 1/6$. 

The enhancement of $\Omega_\text{GW}$ occurs during the kinaton domination regime in which the energy density is dominated by the kinetic energy of a scalar. Then the energy density of GW redshifts as $\propto a^{-4}$ and the energy density of the Universe redshifts as $\propto a^{-6}$ due to the kinaton domination. Therefore $\Omega_\text{GW} \propto a^2$. The period of enhancement is stronger if the era of the kinaton domination last longer, which is the case if the initial ratio of radiation to inflaton is smaller. This ratio can be decreased either by decreasing the scale of inflation or by taking $\xi$ to be close to the conformal value of $\xi$. In the first case one would also decrease the hierarchy between the scale of inflation and BBN, which would give less time for the kinaton domination, and there is not enough room for the GW enhancement.

In our model, the scale of inflation is close to the Starobinsky model. In particular $H_{\rm inf}$ is close to
\begin{equation}
H_{\rm inf} \sim 1.3 \times 10^{13}\GeV,
	\label{highest inflation scale}
\end{equation}
which is the case already analyzed in the Ref. \cite{Artymowski:2017pua}.  The only way to elongate the kinaton domination phase (which leads to the longer period of the primordial GW enhancement) is to make the gravitational reheating even less efficient than usual by tuning $\xi\simeq 1/6$. As shown in Fig.~6 in Ref.~\cite{Artymowski:2017pua}, one can tune the value of $\xi$ for almost any scale of dark inflation and still obtain the enhancement of $\Omega_\text{GW}$. Let us stress that the enhancement of GW is not compulsory for dark inflation's existence or consistency with the experimental data. Without fine tuning of $\xi$ one can still generate enough $e$-folds and correct forms of primordial power spectra. We are only pointing out a possible characteristic signal of the model, observation of which would strongly indicate the existence of dark inflation. We note that all the other realization of the dark inflation with the gravitational reheating with $w=1$ also give the same enhancement of the primordial GW.

As shown in Fig.~6 in Ref.~\cite{Artymowski:2017pua} the value of the contrast function of primordial gravitational waves $\Omega_\text{GW}$ enables future detection for $\textsf{N}_{\rm eff} \equiv \textsf{N} (1-6\xi)^2 \simeq 10^{-4}$, which gives $\xi \simeq \frac{1}{6}(1-10^{-2}/\sqrt{\textsf{N}})$. The same plot proves that the result satisfies the BBN constraint on the primordial gravitational waves, which is the upper bound for the allowed scale of inflation. For such a finely tuned value of $\xi$, the characteristic signal predicted by our model could be measured by DECIGO and BBO \cite{Yagi:2011wg}. 

Since the inflaton in our model belongs to the dark sector, it can naturally be a source of dark matter~\cite{Ema:2018ucl}. Furthermore, the dark inflation can also be responsible for the change of the DM relic density abundance, due to the long period of the inflaton domination after inflation. Such a non-standard thermal history of the Universe has a strong influence on the freeze-out scale of any produced particles, which is defined by the relation $\Gamma = H$, where $\Gamma$ is a decay width of a given particle. In our case the inflaton can dominate the Universe at the freeze out scale, which increases the value of the Hubble parameter for given temperature. For more details on this issue see Refs \cite{Beniwal:2017eik,Redmond:2017tja,Pallis:2005bb}. The same apply to e.g. leptogenesis. This issue was discussed in \cite{Dutta:2018zkg}, where authors show that having a kinaton domination era may significantly decrease the scale of leptogenesis.

\section{Conclusions} \label{sec:conclusions}

In this paper we present a concrete model of a hill-climbing dark inflation with kinaton domination phase. Unlike in the most scalar-tensor theories, the non-minimal coupling to the gravity increases the strength of the gravitational interaction, leading to the inflation and up-hill evolution of the Jordan-frame field. In Sec.~\ref{sec:HCinflation} we have discussed the background evolution of the inflaton. We have proven that the Einstein-frame potential has an inflationary plateau and that inflation can be obtained even for kinetic initial conditions. We have also shown that the lowest scale of inflation should be obtained if mass scales in the Jordan frame potential, $M_V$, and in the function of non-minimal coupling to gravity, $M_F$, are equal to each other. 

In Sec.~\ref{sec:reheating} we have demonstrated that the gravitational reheating can successfully be implemented in our model. Even if the inflaton is decoupled from any other fields, one can produce gravitationally sufficient amount of the relativistic particles at the end of inflation in order to satisfy the BBN constrains on radiation domination at the MeV scale. Here a small amount of radiation eventually dominates the Universe due to the kinetic term domination after inflation. 

In Sec. \ref{sec:inhomogeneities} we present the analysis of primordial inhomogeneities generated during inflation. We have shown that our model is in general consistent with the data. The results weakly depend on $M_F$ and strongly on the ratio  $M_F/M_V$. We have also shown that the dark inflation from our model may be responsible for enhancing the primordial gravitational waves, which could be measured by BBO/DECIGO experiments. Such an enhancement remains consistent with the BBN upper bound of the scale of inflation. Nevertheless it requires certain fine-tuning. Produced scalars should have a non-minimal coupling to gravity, which deviates from the conformal factor at the level of 1\%.                                                                

\section*{Acknowledgements}

We thank Misao Sasaki for useful comments. MA was supported by the Iuventus Plus grant No. 0290/IP3/2016/74 from the Polish Ministry of Science and Higher Education. ZL was supported by the Polish NCN grant DEC-2012/04/A/ST2/00099

\appendix
\section*{Appendix}
\section{Generalization 1}

The model presented in this work can be easily generalized. Let us define the Jordan frame scalar potential and the function of non-minimal coupling as
\begin{equation}
V_\text{J} =  {\lambda\over\Lambda^{2n-4}}\phi_\text{J}^{2n} e^{-(\phi_\text{J}/M_V)^p} \ , \qquad F = 1 - e^{-(\phi_\text{J}/M_F)^n} \, , \label{eq:generalisation}
\end{equation}
where $n$ and $p$ are some natural numbers. For $n=p=1$ with $\lambda\Lambda^2=m^2/2$, one restores Eq.~(\ref{eq:Jordan framepotential}). Alike in the model from Eq.~(\ref{eq:Jordan framepotential}) we want to secure the constant value of the Einstein frame potential in the small field limit. Indeed, for $\phi_\text{J} \ll M_V, M_F$ one finds $F \propto \phi_\text{J}^n$, which gives $V_\text{E} = V_\text{J}/F^2 \propto \phi_\text{J}^{2n}/(\phi_\text{J}^n)^2\sim const$. More detailed calculations show that in the small field limit one finds
\begin{equation}
V_\text{E} \simeq {\lambda\over\Lambda^{2n-4}} M_F^{2n} \left(1-\left(\frac{\phi_\text{J} }{M_V}\right)^p+\frac{1}{2}\left(\frac{\phi_\text{J}}{M_V}\right)^{2p}\right)\left(1+\left(\frac{\phi_\text{J} }{M_F}\right)^n+\frac{5}{12}\left(\frac{\phi_\text{J} }{M_F}\right)^{2n}\right) \, . \label{eq:VEgen}
\end{equation}
The Einstein frame field in the small field limit is defined by
\begin{equation}
\phi_\text{E} \simeq \sqrt{\frac{3}{2}}\log F \qquad \Rightarrow \qquad \left(\frac{\phi_\text{J}}{M_F}\right)^n \simeq e^{\sqrt{2/3}\, \phi_\text{E}} \, . \label{eq:Einstein framefield2}
\end{equation}
In order to compare this generalization with the original model from Eq.~(\ref{eq:Jordan framepotential}) let us assume that $p = n$. Then, in the $M_V \neq M_F$ one finds
\begin{equation}
V_\text{E} \simeq {\lambda M_F^{2n}\over\Lambda^{2n-4}}\left(1-\frac{M_F^n-M_V^n}{M_V^n}\left(\frac{\phi_\text{J}}{M_F}\right)^n\right) \simeq {\lambda M_F^{2n}\over\Lambda^{2n-4}}\left(1-\frac{M_F^n-M_V^n}{M_V^n} e^{\sqrt{2/3}\, \phi_\text{E}}\right)\, .
\end{equation}
This result clearly restores (\ref{eq:V2}) int he $n=1$ limit. Again, one requires $M_F > M_V$ in order to secure the growth of $\phi_\text{J}$. In the simplest case of $M_F = M_V$ and $n = p$ Eq.~(\ref{eq:VEgen}) simplifies to 
\begin{equation}
V_\text{E} \simeq {\lambda\over\Lambda^{2n-4}} M_V^{2n} \left(1-\frac{1}{12}\left(\frac{\phi_\text{J}}{M_V}\right)^{2n}\right) \simeq {\lambda\over\Lambda^{2n-4}} M_V^{2n} \left(1-\frac{1}{12}e^{2 \sqrt{2/3}\, \phi_\text{E}} \right) \, ,
\end{equation}
which fully restores Eq.~(\ref{eq:Vequal}). Thus, the $M_V \to M_F$ limit appears to be a common solution for all of the values of $n$.

\section{Generalization 2}

Another possible generalization of the model is
\begin{equation}
V_\text{J} = {\lambda\over\Lambda^{2n-4}} \phi_\text{J}^{2n} e^{-\frac{\phi_\text{J}}{M_V}} \ , \qquad F = \left(1 - e^{-\frac{\phi_\text{J}}{n\, M_F}}\right)^n \, . \label{eq:generalisation2}
\end{equation}
The $n=1$ case corresponds to the model in the main text with $\lambda\Lambda^2=m^2/2$.
The fundamental difference with the (\ref{eq:generalisation}) is that for even $n$ one finds $F \geq 0$ for all $\phi_\text{J}$ and therefore one avoids the repulsive gravity regime in the theory. In the small field regime one finds
\begin{equation}
V_\text{E} \simeq  {\lambda\over\Lambda^{2n-4}} \left(n\, M_F\right)^{2n} \left(1-\frac{M_F- M_V}{M_F M_V}\phi_\text{J}\right)\, ,
\end{equation}
which reconstructs the result of Eq. (\ref{eq:V1}) . In order to obtain a field going up-hill one requires $M_F > M_V$. For $M_F = M_V$ one finds
\begin{equation}
V_\text{E} \simeq  {\lambda\over\Lambda^{2n-4}} \left(n\, M_V\right)^{2n} \left(1-\frac{1}{12 n M_V^2}\phi_\text{J}^2\right)\, .
\end{equation}
Again, the result is almost identical with Eq. (\ref{eq:V2}). The significant difference comparing to the original model and to the first generalization is the form of the Einstein frame potential in the $M_F \to M_V$ limit. In the case of Eq. (\ref{eq:generalisation2}) one obtains
\begin{equation}
V_\text{E} \simeq {\lambda\over\Lambda^{2n-4}} \left(n\, M_V\right)^{2n} \left(1-\frac{n}{12}e^{2 \sqrt{\frac{2}{3n}}\, \phi_\text{E}} \right) \, ,
\end{equation}
which does not create an attractor for all values of $n$.

\end{document}